\documentclass[a4paper,11pt,reqno]{amsart}

\usepackage{amssymb, amscd}
\usepackage{latexsym,epsfig}
\usepackage[all]{xy}
\numberwithin{equation}{section}

\usepackage[T1]{fontenc}

\def\today{\ifcase\month\or Jan\or Febr\or  Mar\or  Apr\or May\or Jun\or  Jul\or Aug\or  Sep\or  Oct\or Nov\or  Dec\or\fi \space\number\day, \number\year}

\newcommand{\CC}{\mathbb C}

\newcommand{\PP}{\mathbb P}

\newcommand{\RR}{\mathbb R}

\newcommand{\ZZ}{\mathbb Z}

\newcommand{\Sp}{{\mathrm{Sp}}}
\newcommand{\SL}{{\mathrm{SL}}}
\newcommand{\GL}{{\mathrm{GL}}}

\newcommand{\IM}{{\mathrm{Im}}}

\newtheorem{theorem}{Theorem}[section]

\newtheorem{definition-lemma}[theorem]{Definition-Lemma}

\theoremstyle{definition}
\newtheorem{definition}[theorem]{Definition}
\newtheorem{example}[theorem]{Example}

\theoremstyle{remark}
\newtheorem{remark}[theorem]{Remark}

\begin{document}

\title{Dispersionless Hirota equations and the genus 3 hyperelliptic divisor}

\date\today

\author{Fabien Cl\'ery}
\address{Department of Mathematical Sciences,
Loughborough University,
United Kingdom}
\email{cleryfabien@gmail.com}

\author{Evgeny V. Ferapontov}
\address{Department of Mathematical Sciences,
Loughborough University,
United Kingdom}
\email{E.V.Ferapontov@lboro.ac.uk}

\subjclass{11F46, 14J15, 14K25, 37K10, 37K20, 37K25,  53A30, 53B25, 53B50}


\begin{abstract}
Equations of dispersionless Hirota type
\[
F(u_{x_ix_j})=0
\]
have been thoroughly investigated in mathematical physics and differential geometry. 
It is known that the parameter space of integrable Hirota type  equations in 3D is 21-dimensional, and 
that the action of the natural equivalence group $\Sp(6,\RR)$ on the parameter space has an open orbit. 
However the structure of the generic equation corresponding to the open orbit  remained elusive. 
Here we prove that the generic 3D Hirota equation is given by the remarkable formula
\[
 \vartheta_m(\tau)=0, \qquad \tau=i\ \text{Hess}(u)
\]
where $\vartheta_m$ is any genus 3 theta constant with  even characteristics and $\text{Hess}(u)$ is the $3\times 3$ Hessian matrix of a (real-valued) function $u(x_1, x_2, x_3)$. Thus, generic Hirota equation coincides with the equation of the genus 3 hyperelliptic divisor (to be precise, its intersection with the imaginary part of the Siegel upper half space $\mathfrak{H}_3$). 
The rich geometry of integrable Hirota type equations sheds new light on local differential geometry 
of the genus 3  hyperelliptic divisor, in particular, the integrability conditions can be viewed as local differential-geometric constraints that characterise the hyperelliptic divisor uniquely modulo $\Sp(6,\CC)$-equivalence.
\end{abstract}


\maketitle
\centerline{\it Dedicated to the memory of Professor Boris Dubrovin}


\section{\textbf{Introduction}}

A 3D dispersionless Hirota type equation is a second-order PDE of the form 

\begin{equation}\label{main}
F(u_{x_ix_j})=0
\end{equation}
where $u$ is a function of 3 independent variables. Equations of type (\ref{main})  
arise in numerous applications in non-linear physics, general relativity, differential geometry, 
theory of integrable systems, and complex analysis. For instance, 
 dispersionless Kadomtsev-Petviashvili (dKP) equation
\begin{equation}\label{dKP}
u_{x_1x_3}-\frac{1}{2}u_{x_1x_1}^2-u_{x_2x_2}=0
\end{equation} 
arises in  non-linear acoustics \cite{KZ}, the theory of Einstein-Weyl structures \cite{D} and 
in the context of conformal maps of simply-connected domains \cite{Wiegmann, Zab, Marshakov}. 
The Boyer-Finley (BF) equation
\[
u_{x_1x_1}+u_{x_2x_2}-e^{u_{x_3x_3}}=0
\]
describes a class of self-dual Ricci-flat $4$-manifolds \cite{BF}. Let us review briefly
some of the known properties of {\it integrable} Hirota type equations
(see Section \ref{sec:I} for a summary of four equivalent approaches to dispersionless integrability in 3D):

\begin{itemize}
\item The class of integrable equations (\ref{main}) is invariant under the  symplectic group $\Sp(6,\RR)$  
via its standard action
\begin{equation}\label{Sp}
U \mapsto (AU+B)(CU+D)^{-1}.
\end{equation}
Here $U=\text{Hess}(u)$ is the Hessian matrix of the function $u$ \cite{FHK}. 
The classification of integrable equations is performed modulo equivalence  (\ref{Sp}).

\item The parameter space of  integrable equations (\ref{main})  is 21-dimensional (see theorem 1 of \cite{FHK}). 
Furthermore, the action of the equivalence group  $\Sp(6,\RR)$ on the parameter space  is locally free. 
The fact that the group $\Sp(6,\RR)$ is 21-dimensional  implies the existence of a generic Hirota equation 
generating an open 21-dimensional $\Sp(6,\RR)$-orbit. Any representative of the open orbit will be called a {\it Hirota master-equation}. The master-equation has no {\it continuous} symmetries 
from the equivalence group. Note that both dKP and BF equations have non-trivial symmetry groups and  
generate singular orbits of lower dimensions (14 and 15, respectively).

\item Geometrically, Hirota type equation (\ref{main}) can be viewed as the defining equation of a 5-dimensional 
hypersurface $M^5$ in the 6-dimensional Lagrangian Grassmannian $\Lambda^6$ 
(locally identified with $3\times 3$ symmetric matrices). Integrability of equation (\ref{main}) 
can be interpreted as integrability of a $\GL(2, \RR)$ geometry induced on $M^5$ (see \cite{Smith} and section 4 of \cite{FHK}). 

\end{itemize}

Although the structure of the Hirota master-equation has remained elusive, 
formula (\ref{Sp}) suggests that this equation should possess non-trivial modular properties. 
In this direction let us mention yet another integrable example,

\begin{equation}\label{Chazy}
u_{x_3x_3} -\frac{u_{x_1x_2}}{u_{x_1x_3}}-\frac{1}{6}h(u_{x_1x_1})u_{x_1x_3}^{2}=0
\end{equation}
which was discussed in  \cite{MaksEgor1,MaksEgor2}. Here  the function $h$ satisfies the Chazy equation
\[
h^{\prime \prime \prime }+2hh^{\prime \prime }-3(h^{\prime}) ^2=0
\]
whose general solution can be expressed in terms of the  Eisenstein series of weight 2 on the 
modular group $\SL(2,\ZZ)$:  $h(t)=e_2(i t/\pi)$ where
\begin{equation}\label{Eis2}
e_2(\tau)=1-24\sum_{n=1}^{\infty} \sigma_1(n)e^{2\pi in\tau}=1-24\,q-72\,q^2-96\, q^3+\dots
\quad
\text{where}
\quad
q=e^{2\pi i \tau}.
\end{equation}
Note that $h(t)$ is real for $t>0$. This was one of the first examples where modular forms explicitly 
occurred as {\it coefficients} rather than as {\it solutions} of integrable systems, see also \cite{Krich+Zabr, FK, Buch}; note that  PDE  (\ref{Chazy}) admits non-trivial continuous symmetries and therefore does not generate an open orbit. 
We refer to section \ref{sec:ex} for further `modular' examples.

Before giving our main theorem, let us recall few facts  about theta constants. 
All these facts can be found in \cite{I2}. 
Let $\mathfrak{H}_3$ be the Siegel upper half space of genus 3 i.e.
\[
\mathfrak{H}_3=
\left\{
\tau \in \text{Mat}(3\times3,\CC)\, | \, \tau=\tau^t, \IM(\tau)>0 
\right\}.
\]
Theta constants with characteristics are defined by
\[
\vartheta_m(\tau)=\vartheta_
{
\left[
\begin{smallmatrix}
\mu\\
\nu
\end{smallmatrix}
\right]
}
(\tau)=
\sum_{n\in \mathbb{Z}^3}
e^{i\pi (n+\mu/2)\left(\tau(n+\mu/2)^t+\nu^t\right)}
\]
where $\mu, \nu \in \{0, 1\}^3$ and $\tau\in \mathfrak{H}_3$. The characteristic 
$m=\left[
\begin{smallmatrix}
\mu\\
\nu
\end{smallmatrix}
\right]$
is called even if $\mu \nu^t$ is even. In genus 3, there are 36 such characteristics and they give
rise to 36 theta constants which are modular forms of weight one-half with a miltiplier system (see \cite{Glass}, theorem 1.1) on the so-called Igusa group $\Gamma_3(4, 8)$ defined as
$$
\Gamma_3(4, 8)=\{ M=\left(\begin{array}{cc}
a&b\\
c&d
\end{array}\right)
\in \Sp(6,\ZZ): \  M\equiv I_6 \ {\rm mod}\ 4,\ {\rm diag}(a^tb)\equiv {\rm diag}(c^td)\equiv 0 \ {\rm mod}\ 8 \}.
$$
Recall that there is an action of the group $\Sp(6,\ZZ)$ on the set of even characteristics which is transitive. 
The product of  36 theta constants with even characteristics equals a cusp form of weight 18 on the full 
modular group $\Sp(6,\ZZ)$, this form is classically denoted by $\chi_{18}$ (see \cite{Igusa}). It is known that in genus 3
the vanishing of even theta constants characterises the hyperelliptic divisor which consists of 36 irreducible isomorphic components. Note that $\vartheta_m(\tau)$ is real when $\tau$ is purely imaginary.

Our main result is the following explicit formula for the master-equation:

\begin{theorem}\label{T1}
The master-equation is given by the formula 
\begin{equation}\label{theta}
 \vartheta_m(\tau)=0, \qquad \tau=i\ \text{Hess}(u)
\end{equation}
where $\vartheta_m$ is any genus 3 theta constant with even characteristic $m$ and $\text{Hess}(u)$ is the Hessian matrix of the function $u$. Equation (\ref{theta}) defines a hypersurface $M^5$ in the Lagrangian Grassmannian $\Lambda^6$ (we  assume $\text{Hess}(u)$  to be positive definite).  Embedding the positive cone of $\Lambda^6$ into the purely imaginary locus of the Siegel upper half space $\mathfrak{H}_3$ via multiplication by $i$, we can thus characterise $M^5$ as the intersection of the purely imaginary locus with the genus 3 hyperelliptic divisor. 
\end{theorem}

\begin{remark}
Note that although there are 36 even theta constants in genus 3, the corresponding equations (\ref{theta}) 
are all equivalent (over $\CC$) due to the transitivity of the $\Sp(6,\ZZ)$-action on the set of even theta constants.
Using the cusp form $\chi_{18}$, the equations (\ref{theta}) can be compactly represented as
\[
\chi_{18}(i\ \text{Hess}(u))=0.
\]
\end{remark}




Theorem \ref{T1} will be proved in section \ref{sec:OS} by uncovering geometry behind  the Odesskii-Sokolov 
construction \cite{Odesskii2} that parametrises broad classes of dispersionless integrable systems 
via generalised hypergeometric functions.

There exist several  approaches to dispersionless integrability in 3D. Based on seemingly different 
ideas they however lead to equivalent classification results. These approaches are:

\begin{itemize}

\item The method of hydrodynamic reductions based on the requirement that equation (\ref{main}) has 
`sufficiently many' special multiphase solutions. 

\item Integrability {\it on equation}, meaning that the associated hypersurface $M^5\subset \Lambda^6$ 
carries an integrable $\GL(2,\RR)$ geometry.

\item Integrability {\it on solutions}, based on the requirement that the characteristic variety of equation (\ref{main}) 
defines Einstein-Weyl geometry on every solution.

\item Integrability via a dispersionless Lax representation.
\end{itemize}

The above properties of the master-equation shed new light on local differential geometry of the genus 3 hyperelliptic divisor, 
see sections \ref{sec:HR}--\ref{sec:L}.

\section{\textbf{Four equivalent approaches to dispersionless integrability}}
\label{sec:I}

In this section we  summarise the existing approaches  to dispersionless integrability in 3D.

\subsection{Integrability via hydrodynamic reductions}\
\label{sec:HR}

The method of hydrodynamic reductions applies to quasilinear systems of the form
 \begin{equation}
A ({\bf v}){\bf v}_{x_1}+B({\bf v}){\bf v}_{x_2}+C({\bf v}){\bf v}_{x_3}=0
\label{quasi1}
 \end{equation}
where ${\bf v}=(v^1, ..., v^m)^t$ is an $m$-component column vector of the dependent variables and $A, B, C$ 
are $l\times m$ matrices, $l\geq m$. Note that Hirota type equation  (\ref{main}) can be brought to  
quasilinear form  (\ref{quasi1}) by representing it in evolutionary form,
\begin{equation}\label{evol}
u_{x_1x_1}=f(u_{x_1x_2}, u_{x_1x_3}, u_{x_2x_2}, u_{x_2x_3}, u_{x_3x_3}),
\end{equation}
choosing the arguments of  $f$  as the new dependent variables ${\bf v}$ and writing out all possible consistency 
conditions among them. This results in the quasilinear representation  (\ref{quasi1}) with  $m=5,  \ l=8$. 
Applied to system (\ref{quasi1}) the method of hydrodynamic reductions consists of seeking multiphase solutions in the form
 \begin{equation*}
{\bf v}={\bf v}(R^1, ..., R^N)
 \end{equation*}
where the phases (Riemann invariants) $R^i(x_1, x_2, x_3)$, whose number $N$ is allowed to be arbitrary, are required 
to satisfy a pair of commuting $(1+1)$-dimensional systems
 \begin{equation}\label{R}
R^i_{x_2}=\mu^i(R) R^i_{x_1},\quad R^i_{x_3}=\lambda^i(R) R^i_{x_1}
 \end{equation}
known as systems of hydrodynamic type \cite{DN, Tsar}. The corresponding characteristic speeds $\mu^i, \lambda^i$ 
must satisfy the commutativity conditions \cite{Tsar}
\begin{equation*}
\frac{\partial_j\mu^i}{\mu^j-\mu^i}=\frac{\partial_j\lambda^i}{\lambda^j-\lambda^i},
 \end{equation*}
where $i, j\in\{1, \dots, N\}, \ i\ne j, \  \partial_j=\partial_{ R^j}$.
Equations (\ref{R}) are said to define an $N$-component  hydrodynamic reduction of  system (\ref{quasi1}). 
The following definition was proposed in \cite{FK}, see also  \cite{GibTsar1}:

\begin{definition} System (\ref{quasi1}) is said to be {\it integrable} if, for every $N$, 
it admits infinitely many $N$-component hydrodynamic reductions parametrised by $2N$ arbitrary functions of one variable.
\end{definition}

This requirement imposes strong  constraints (integrability conditions) on the matrix entries of  $A, B$ and $C$. 
Applied to equation (\ref{evol}), the method of hydrodynamic reductions leads to an $\Sp(6,\CC)$-invariant 
set of  differential constraints for the function $f$ expressing all third-order partial derivatives of $f$ in terms 
of its first and second-order partial derivatives (35 relations  that are rational in  the partial derivatives of $f$). 
The integrability conditions were first derived in \cite{FHK}. Their involutivity implies that the parameter space of 
integrable Hirota type equations is 21-dimensional \cite{FHK}. 
 
Note that  (\ref{evol}) is, locally,   the equation of the graph of the corresponding hypersurface 
 $M^5\subset \Lambda^6$. Our main result (theorem \ref{T1}) states that  Hirota master-equation coincides with the equation of the genus 3 
 hyperelliptic  divisor. Thus, the integrability conditions can be viewed as  local differential constraints that 
 characterise the hyperelliptic divisor uniquely modulo $\Sp(6,\CC)$-equivalence.

\subsection{Integrability on equation:  integrable $\GL(2,\RR)$ geometry}\
\label{sec:GL}

The Lagrangian Grassmannian $\Lambda^6$ (locally parametrised by  $3\times 3$ symmetric matrices)
 carries a flat generalised conformal structure defined by the family of degree 4 Veronese cones 
in  $T\Lambda^6$ (identified with rank 1 symmetric matrices). Let $M^5$ be a hypersurface in $\Lambda^6$. 
Taking a point $s\in M^5$ and intersecting the tangent space $T_sM^5$ with the 
Veronese cone in $T_s\Lambda^6$ one obtains a rational normal cone of degree 4 in $T_sM^5$. 
On projectivisation, this results in a family of rational normal curves $\gamma$ of degree 4 in $\PP TM^5$. 
This structure is known as  $\GL(2,\RR)$ geometry on $M^5$.  

\medskip

\begin{definition}\

A {\it bisecant plane} in $TM^5$ is a plane whose projectivisation is a bisecant line of 
$\gamma$. 

A {\it bisecant surface} is a 2-dimensional submanifold $\Sigma^2\subset M^5$ whose  tangent planes are bisecant. 

A {\it trisecant space} in $TM^5$ is a 3-dimensional subspace whose projectivisation is a trisecant plane of $\gamma$. 

A {\it trisecant $3$-fold} is a 3-dimensional submanifold $\Sigma^3\subset M^5$ whose  tangent spaces are trisecant.  
\end{definition}

To be more precise, we will need {\it holonomic} trisecant $3$-folds which can be defined as follows. Note that each tangent space $T\Sigma^3$ carries 3 distinguished directions, namely those corresponding to the 3 points of intersection of $\PP T\Sigma^3$ with $\gamma$.
These directions define a net on $\Sigma^3$ which we require  to be  holonomic i.e. a coordinate net.

It turns out that  bisecant surfaces and holonomic trisecant $3$-folds of a hypersurface $M^5$  
correspond to 2- and 3-component hydrodynamic reductions of the associated  Hirota type equation. 
Furthermore, the hypersurface $M^5$ corresponds to an integrable equation  if and only if it has infinitely many 
holonomic trisecant $3$-folds  parametrised by 3 arbitrary functions of one variable \cite{FHK}. 
Thus the existence of holonomic trisecant $3$-folds is a geometric interpretation of the integrability property. 
The corresponding integrable $\GL(2,\RR)$ geometries were thoroughly investigated in \cite{Smith}. 

 Thus,  theorem \ref{T1} implies that the genus 3 hyperelliptic divisor carries  integrable $\GL(2,\RR)$ geometry.

\subsection{Integrability on solutions:  Einstein-Weyl geometry}\
\label{sec:EW}

Any solution of equation (\ref{main}) carries a family of  characteristic cones
 \[
 \frac{\partial F}{\partial u_{x_ix_j}}\ p_{i} p_{j}=0,
 \]
assumed to be non-degenerate. 
The inverse matrix of the associated quadratic form gives  rise to the conformal structure
$[g]=g_{ij} dx_{i} dx_{j}$ which depends on a solution due to non-linearity of (\ref{main}).  
It was shown in theorem 7 of \cite{FerKrug}  that  integrability of  equation (\ref{main})  is equivalent to the 
requirement that  conformal structure $[g]$ is Einstein-Weyl on every solution. 
Recall that an Einstein-Weyl structure consists of a symmetric connection $\mathbb{D}$ 
and a conformal structure $[g]$  such that

\begin{itemize}

\item connection $\mathbb {D}$ preserves the conformal class: $\mathbb{D}[g]=0$,

\item trace-free part of the symmetrised Ricci tensor of $\mathbb {D}$ vanishes.

\end{itemize}
In local coordinates, 
\begin{equation}\label{EW}
\mathbb{D}_kg_{ij}=\omega_k g_{ij}, 
\quad 
R_{(ij)}=\rho g_{ij},
\end{equation}
where $\omega=\omega_kdx_k$ is a covector, $R_{(ij)}$ is the symmetrised Ricci tensor 
of $\mathbb{D}$,  and $\rho$ is some function.  It was shown in section 5 of \cite{FerKrug} that 
the covector $\omega$ can be  expressed in terms of  $g$ by the  universal explicit formula
\begin{equation}\label{omega}
\omega_k=2g_{kj}\mathcal{D}_{x_s}(g^{js})+\mathcal{D}_{x_k}(\ln\det g_{ij})
\end{equation}
where $\mathcal{D}_{x_k}$ denotes the total derivative with respect to $x_k$. 
We recall that in 3D the Einstein-Weyl equations  (\ref{EW}) are  integrable by   
twistor-theoretic methods \cite{Hitchin}. 
Thus solutions of integrable PDEs carry `integrable' conformal geometry.  
Combining results of sections 2.2 and 2.3 with theorem \ref{T1} we can conclude that every trisecant 3-fold of the hyperelliptic divisor 
carries Einstein-Weyl geometry.
 
Cartan proved (see \cite{Cartan}) that a pair $(\mathbb{D}, [g])$ defined on a 3-dimensional manifold satisfies  
Einstein-Weyl equations (\ref{EW}) if and only if there exists a 2-parameter family of
surfaces  which are totally geodesic with respect to the connection $\mathbb{D}$ and null with respect 
to the conformal structure $[g]$.  Such surfaces come from the associated dispersionless 
Lax pairs \cite{D, FerKrug, CalKrug}.

\subsection{Integrability via dispersionless Lax representation}\
{\label{sec:L}

A pair of Hamilton-Jacobi type equations for an auxiliary function $S$,
\begin{gather}
\begin{aligned} \label{S1}
S_{x_2}& \ =f(S_{x_1},u_{x_ix_j}), \\
 S_{x_3}& \ =g(S_{x_1},u_{x_ix_j}),
\end{aligned}
\end{gather}
is said to define a dispersionless Lax representation of equation (\ref{main}) if (\ref{S1})  
is compatible modulo (\ref{main}). Lax pairs of this form appeared in \cite{Zakharov} as 
dispersionless limits of  Lax pairs of integrable soliton equations. 

\begin{example} Dispersionless Lax pair of the dKP equation (\ref{dKP}) has the form
\begin{gather}
\begin{aligned} \label{S3}
S_{x_2}& \ =\frac{1}{2}S_{x_1}^2+u_{x_1x_1}, \\
 S_{x_3}& \ =\frac{1}{3}S_{x_1}^3+u_{x_1x_1}S_{x_1}+u_{x_1x_2}.
\end{aligned}
\end{gather}
\end{example}

\begin{example} Dispersionless Lax pair of  equation (\ref{Chazy}) has the form
\begin{gather*}
\begin{aligned}
S_{x_2}&\, =f(S_{x_1},u_{x_1x_1})u_{x_1x_2}+\frac{1}{3}g(S_{x_1},u_{x_1x_1})u_{x_1x_3}^3, \\
S_{x_3}& \,=f(S_{x_1},u_{x_1x_1})u_{x_1x_3}.
\end{aligned}
\end{gather*}
Here the functions $f$ and $g$ are defined by (we set $S_{x_1}=x, \ u_{x_1x_1}=t$):
\[
g=f_t+ff_x,
\quad
f=-2(\ln v)_x
\]
where the function $v$ is the Jacobi theta series $\vartheta_{00}$, see (\ref{JacobiTheta}) for the definition,
evaluated at $(\tau,z)=(\frac{i t}{\pi},\frac{x}{2\pi})$:
\[
v(x,t)=\vartheta_{00}(i t/\pi,x/2\pi)=1+2\sum_{n=1}^{\infty}e^{-n^2t}\cos(nx).
\]
\end{example}

In some cases it is more convenient to deal with parametric Lax pairs,
\begin{equation}\label{S2}
S_{x_1}=f(p,u_{x_ix_j}),
\quad
S_{x_2}=g(p,u_{x_ix_j}), 
\quad
S_{x_3}=h(p,u_{x_ix_j}), 
\end{equation}
where $p$ is an auxiliary parameter. Such Lax pairs were used in the construction of the universal 
Whitham hierarchy \cite{Kr}. For example, parametric form of the dKP Lax pair (\ref{S3}) is
\[
S_{x_1}=p, ~~~ S_{x_2}=\frac{1}{2}p^2+u_{x_1x_1}, ~~~ S_{x_3}=\frac{1}{3}p^3+u_{x_1x_1}p+u_{x_1x_2}.
\]
Note that the compatibility condition of equations (\ref{S2}) is 
\[
f_p(g_{x_3}-h_{x_2})+g_p(h_{x_1}-f_{x_3})+h_p(f_{x_2}-g_{x_1})=0.
\]
It follows from \cite{FerKrug, CalKrug} that Hirota type equation (\ref{main}) is integrable if and only if 
it has a dispersionless Lax representation (satisfying a suitable non-degeneracy condition).

\section{\textbf{Hirota master-equation via the Odesskii-Sokolov construction: Proof of theorem \ref{T1}}}\
\label{sec:OS}

It was proved in  \cite{Odesskii2} that a generic integrable Hirota type equation (\ref{main})  can be parametrised 
by generalised hypergeometric functions. Here we briefly summarise the construction. 
Consider the generalised hypergeometric system of Appell's type,
\begin{gather}
\begin{aligned} \label{hyper}
 \frac{\partial^2 h}{\partial v_i \partial v_j}=&
 \frac{s_i}{v_i-v_j} \frac{\partial h}{\partial v_j}
 +\frac{s_j}{v_j-v_i} \frac{\partial h}{\partial v_i}, \quad i \ne j, \\
\frac{\partial^2 h}{\partial {v_i}^2}=&-\sigma \frac{s_i}{v_i (v_i-1) }\, h+
\frac{s_i}{v_i (v_i-1)} \sum_{j\ne i}^n \frac{v_j(v_j-1)}{v_j-v_i}
\frac{\partial h}{\partial v^j}\\
&+ \Big(\sum_{j\ne i}^n \frac{s_j}{v_i-v_j}+
\frac{s_i+s_{n+1}}{v_i}+ \frac{s_i+s_{n+2}}{v_i-1}\Big)
\frac{\partial h}{\partial v_i}.
\end{aligned}
\end{gather}
Here   $s_1,...,s_{n+2}$ are arbitrary constants (which we will assume to be rational), $\sigma=1+s_1+\dots +s_{n+2}$, 
and $h$ is a function of $n$ variables $v_1, \dots, v_n$. 
This system is involutive and has $n+1$ linearly independent solutions   
known as  generalised hypergeometric functions \cite{gel, Odesskii2}. 
Introducing the differential
\[
\omega=t^{s_{n+1}}(t-1)^{s_{n+2}}(t-v_1)^{s_1}\dots (t-v_n)^{s_n}dt,
\]
solutions to system (\ref{hyper}) can be represented in terms of the corresponding periods, 
$
\int_p^q\omega$  where $p, q \in \{0, 1, \infty, v_1, \dots, v_n\}
$. 
This statement can be explicitly found in Mostow \cite{Mostow86} who also noted   that only $n+1$ of these periods are 
linearly independent \cite{Mostow81},  see also  \cite{Deligne, Holzapfel}. 
In low dimensions, analogous observations were made by  Picard in 1883 \cite{Picard}. 
With any generalised hypergeometric system (\ref{hyper}) Odesskii and Sokolov associated a dispersionless 
integrable system in 3D possessing a dispersionless Lax representation \cite{Odesskii2}.

We will need a particular case of the general construction with $n=5$ and  $s_4=s_5=\frac{1}{2}$, $s_1=s_2=s_3=s_6=s_7=-\frac{1}{2}$, see example 5 of \cite{Odesskii2}. 
The first observation of \cite{Odesskii2} is that one can choose a basis of solutions $h_1, h_2, h_3$, 
$g_1, g_2, g_3$ of the corresponding system (\ref{hyper}) such that

\begin{equation}\label{basis}
\begin{array}{c}
\Delta(g_1, h_3, h_1)=\Delta(g_2, h_2, h_3), \\
\Delta(g_1, h_1, h_2)=\Delta(g_3, h_2, h_3), \\
\Delta(g_2, h_1, h_2)=\Delta(g_3, h_3, h_1),
\end{array}
\end{equation}
where we use the notation
\[
\Delta(f_1, f_2, f_3)=\det 
\left(
\begin{smallmatrix}
f_1 & f_2 & f_3\\
f_{1,v_4} & f_{2, v_4} & f_{3, v_4}\\
f_{1, v_5} & f_{2, v_5} & f_{3, v_5}
\end{smallmatrix}
\right).
\]
With these data Odesskii and Sokolov associated a dispersionless Hirota type equation (\ref{main}) 
for an auxiliary function $\mathcal{U}$ depending on $x_1, x_2, x_3$ represented parametrically as 

\begin{equation}\label{pargen1}
\begin{array}{c}
\frac{\Delta(g_1, h_2, h_3)}{\Delta}=\mathcal{U}_{x_1x_1}, \quad \frac{\Delta(g_2, h_3, h_1)}{\Delta}=\mathcal{U}_{x_2x_2}, \quad
\frac{\Delta(g_3, h_1, h_2)}{\Delta}=\mathcal{U}_{x_3x_3}, \\
\ \\
\frac{\Delta(g_1, h_3, h_1)}{\Delta}=\mathcal{U}_{x_1x_2}, \quad \frac{\Delta(g_1, h_1, h_2)}{\Delta}=\mathcal{U}_{x_1x_3}, \quad
\frac{\Delta(g_2, h_1, h_2)}{\Delta}=\mathcal{U}_{x_2x_3},
\end{array} 
\end{equation}
with $\Delta=\Delta(h_1, h_2, h_3)$. 
The required equation results on the elimination of the 5 parameters $v_1, \dots, v_5$ from the 6 relations (\ref{pargen1}).  
This equation was shown to be integrable via a dispersionless Lax representation of type (\ref{S2}).

The geometry behind this construction is as follows.
With the choice of constants $s_i$ specified above the differential $\omega$ takes the form
\[
\omega=\frac{(t-v_4)(t-v_5)}{v}dt,
\quad 
v=\sqrt {t(t-1)(t-v_1)\dots (t-v_5)},
\]
which is a holomorphic differential  on the genus 3 hyperelliptic curve $\mathcal{C}$
\[
v^2=t(t-1)(t-v_1)\dots (t-v_5).
\]
Choosing a system of cycles $a_i, b_j$ on the curve $\mathcal{C}$ with the intersection matrix $a_i\cdot b_j=\delta_{ij}$ 
we  denote by $h_1, h_2, h_3$, $g_1, g_2, g_3$ the corresponding periods of $\omega$ which form 
a basis of solutions of the associated system (\ref{hyper}). 
We will see that this basis automatically satisfies constraints (\ref{basis}) due to the Riemann relations. 
The basis of holomorphic differentials on $\mathcal{C}$ is given by
\[
\omega=\frac{(t-v_4)(t-v_5)}{v}dt,
\quad 
\frac{\partial \omega}{\partial{v_4}}=-\frac{1}{2}\frac{(t-v_5)}{v}dt, 
\quad  
\frac{\partial \omega}{\partial{v_5}}=-\frac{1}{2}\frac{(t-v_4)}{v}dt.
\]
Their periods over $a$ and $b$ cycles are given by the matrix
\[
\left(
\begin{smallmatrix}
h_1 & h_2 & h_3&g_1&g_2&g_3\\
 h_{1, v_4} & h_{2, v_4} & h_{3, v_4}&g_{1, v_4} & g_{2, v_4} & g_{3, v_4}\\
 h_{1, v_5} & h_{2, v_5} & h_{3, v_5}&g_{1, v_5} & g_{2, v_5} & g_{3, v_5}
\end{smallmatrix}
\right).
\]
Multiplying this matrix from the left by the inverse of 
$
\left(
\begin{smallmatrix}
h_1 & h_2 & h_3\\
 h_{1, v_4} & h_{2, v_4} & h_{3, v_4}\\
 h_{1, v_5} & h_{2, v_5} & h_{3, v_5}
\end{smallmatrix}
\right)
$
we get 
\[
\left(
\begin{smallmatrix} 
1 & 0 & 0&\tau_{11}&\tau_{12}&\tau_{13}\\
0 & 1 & 0&\tau_{21} & \tau_{22} & \tau_{23}\\
0 & 0& 1&\tau_{31} & \tau_{32} & \tau_{33}
\end{smallmatrix}
\right)=(I_3\, \tau)
\]
where $\tau$ is the  period matrix of the curve $\mathcal{C}$. Explicitly we have
\begin{equation}\label{pargen11}
\tau=\frac{1}{\Delta}
\left(
\begin{smallmatrix}
\Delta(g_1, h_2, h_3) & \Delta(g_2, h_2, h_3) & \Delta(g_3, h_2, h_3)\\
\Delta(g_1, h_3, h_1) & \Delta(g_2, h_3, h_1) & \Delta(g_3, h_3, h_1)\\
\Delta(g_1, h_1, h_2) & \Delta(g_2, h_1, h_2) & \Delta(g_3, h_1, h_2)
\end{smallmatrix}
\right).
\end{equation}
The symmetry of the period matrix $\tau$ is equivalent to constraints (\ref{basis}). 
Bases satisfying (\ref{basis}) (such that matrix (\ref{pargen11}) has  positive definite imaginary part) are thus in one-to-one correspondence with  canonical systems of cycles on $\mathcal{C}$. 
In particular, all of them are  $Sp(6, \mathbb{R})$-related.
Finally, relations (\ref{pargen1}) take the form $\tau=\text{Hess}(\mathcal{U})$. The elimination of the parameters $v_i$ 
yields a Hirota type equation for $\mathcal{U}$  which by construction is the equation of the genus 3 hyperelliptic divisor,
\[
\vartheta_m(\tau)=\vartheta_m(\text{Hess}(\mathcal{U}))=0,
\]
where $\vartheta_m$ is any even theta constant, a result that goes back to Schottky, 1880, see e.g. \cite{Poor}.  Although there are 36 even theta constants, the corresponding equations are all equivalent due to the transitivity of the  $\Sp(6,\ZZ)$-action on the set of even theta characteristics. Note that if all parameters (branch points) $v_i$ of the hyperelliptic curve are real we can choose a basis of cycles such that  the corresponding period matrix $\tau$ will be purely imaginary. Then $\mathcal{U}=iu$ will also be purely imaginary, and the Hirota equation 
\[
\vartheta_m(i\ \text{Hess(u)})=0
\]
will be real.

The uniqueness (over $\mathbb{C}$) of the Hirota master-equation can be established as follows. Up to a suitable equivalence transformation, every  equation (\ref{main}) can be brought to a form where $\frac{\partial F}{\partial u_{x_1x_1}}\ne 0$. Thus, without any loss of generality, we can work with  the evolutionary equation (\ref{evol}). 
The integrability conditions derived in section 7 of \cite{FHK} express all third-order partial derivatives of $f$ as explicit  rational functions of its first-order and second-order partial derivatives,  symbolically,
\begin{equation}\label{df}
d^3f=P(df, d^2f).
\end{equation}
Equations (\ref{df}) define a rational connected  $21$-dimensional affine variety $X$ in the affine space of third-order jets of $f$.  The algebraic equivalence group $\Sp(6,\CC)$ (of dimension $21$) acts on $X$ in a locally free way (see \cite{FHK}, note that local freeness holds over $\mathbb{C}$ as well), and therefore possesses a Zariski  open orbit. The uniqueness of this orbit follows from the fact that any two Zariski open sets on a connected variety must necessarily intersect.

\section{\textbf{Examples}}
\label{sec:ex}

In this section, we give a few more examples of integrable (non-generic) equations  (\ref{main}) with non-trivial 
modular properties.

\subsection{Equations of the form $u_{x_1x_2} + u_{x_1x_3} f(u_{x_2x_3}, u_{x_3x_3})= 0$}\

This example has already been investigated \cite{Buch, BFT, Fer4}. For the reader's convenience, we
explain it again. In this case, the integrability conditions obtained via the method of hydrodynamic reductions lead 
to the following system of third-order PDEs, we set $b=u_{x_2x_3}$ and $c=u_{x_3x_3}$:
\begin{gather}\label{SPDE}
\begin{aligned}
(ff_{b}-f_{c})f_{bbb}=&\, f_{bb}\,(ff_{bb}+f_{b}^2-f_{bc})=f_{bb}(f\,f_{b}-f_{c})_b,\\
(ff_{b}-f_{c})f_{bbc}=&\, f_{bb}\,(ff_{bc}+f_{b}f_{c}-f_{cc})=f_{bb}(f\,f_{b}-f_{c})_c,\\
(ff_{b}-f_{c})f_{bcc}=&\, 
2\,ff_{bc}^2
-f f_{bb}f_{cc}
-f_{bc}f_{cc}
+f_{b}^2f_{cc}
-2\,f_{b}f_{c}f_{bc}
+2\,f_{c}^2f_{bb},\\
(ff_{b}-f_{c})f_{ccc}=&\,
2\,f^2f_{bc}^2
-2\,f^2f_{bb}f_{cc}
+f f_{bc}f_{cc}
+4\, f f_{b}^2 f_{cc}\\
&-8\,f f_{b} f_{c} f_{bc} 
+4\,f f_{c}^2f_{bb}
-f_{cc}^2
- f_{b}f_{c}f_{cc}
+2\,f_{c}^2 f_{bc}.
\end{aligned}
\end{gather}
We will show that a generic solution of this system is given by the logarithmic derivative of 
any Jacobi theta series with characteristics.

The first two equations lead to the Burgers' equation
$
ff_{b}-f_{c}=\nu f_{bb}
$
where $\nu \in \CC^*$.
Without any loss of generality, we set $\nu=-1$ and using classical results about  Burgers' equation, 
we can write the function $f$ as 
$
f=2\frac{\psi_b}{\psi}
$
where the function $\psi$ satisfies the heat equation:
$
\psi_c=\psi_{bb}.
$
By substituting 
$
f=2\frac{\psi_b}{\psi}
$
in the equations (\ref{SPDE})
and reducing  modulo $\psi_c=\psi_{bb}$, the first two equations
are automatically satisfied while the third one gives the following sixth-order
ordinary differential equation ($\psi^{(i)}$ stands for $\frac{\partial^i \psi}{\partial b^i}$)
\begin{gather}\label{ODE}
\begin{aligned}
&(\psi'''\psi^{(6)}-\psi^{(4)}\psi^{(5)})\psi^4+
(2\psi'''((\psi''')^2-\psi'\psi^{(5)})+3\psi''(\psi''\psi^{(5)}-\psi'\psi^{(6)})\\
&+5\psi^{(4)}(\psi'\psi^{(4)}-\psi''\psi'''))\psi^3+(2(\psi')^3\psi^{(6)}+6(\psi')^2\psi''\psi^{(5)}-10(\psi')^2\psi'''\psi^{(4)}\\
&+2\psi'\psi''\psi'''^2)\psi^2+6(2(\psi')^3\psi'''^2-(\psi')^4\psi^{(5)}-(\psi'\psi'')^2\psi''')\psi\\
&+6(\psi')^3((\psi')^2\psi^{(4)}-2\psi'\psi''\psi'''+(\psi'')^3)=0.
\end{aligned}
\end{gather}
Finally the last equation of (\ref{SPDE}) gives the derivative with respect to $b$ of the latter one. 
We introduce the following Jacobi theta series with characteristics:
\begin{equation}\label{JacobiTheta}
\vartheta_{\alpha\beta}(\tau,z)=
\sum_{n\in \ZZ}
e^{\pi i ((n+\alpha/2)^2\tau+2(n+\alpha/2)(z+\beta/2))}
\end{equation}
where $\tau \in \mathfrak{H}_1=\left\{\tau \in \CC\, | \, \IM(\tau)>0\right\}$, $z \in \CC$ 
and $(\alpha,\beta)\in \RR^2$ (characteristics). 
Recall that each Jacobi theta function satisfies the following PDE (heat equation):
\[
4 i \pi \, \frac{\partial \vartheta_{\alpha\beta}}{\partial \tau}=
\frac{\partial^2 \vartheta_{\alpha\beta}}{\partial z^2}.
\] 
So the function 
$
\psi(b,c)=\vartheta_{\alpha\beta}(\frac{i c}{\pi},\frac{b}{2\pi})
$ 
satisfies the equation 
$
\psi_c=\psi_{bb}.
$
It remains to check that such a function $\psi$ satisfies the equation (\ref{ODE}).
Note that if $h$ is a solution of (\ref{ODE}) then the following two expressions
\[
f_1(b, c)=h(b+g_1(c),c),
\quad
f_2(b, c)=g_2(c)\,e^{b\, g_3(c)}h(b,c)
\]
are also solutions of (\ref{ODE}) for any functions $g_i$; for $f_1$ the proof is obvious
while for $f_2$ it is a short computation. Using the formula
\[
\psi(b,c)=\vartheta_{\alpha\beta}(\frac{i c}{\pi},\frac{b}{2\pi})=
e^{\frac{\pi i}{2}\alpha\beta}
e^{-c\frac{\alpha^2}{4}}
e^{\frac{i\alpha}{2}b}
\vartheta_{00}(\frac{i c}{\pi},\frac{b}{2\pi}+\frac{i c}{\pi}\frac{\alpha}{2}+\frac{\beta}{2}),
\]
it is therefore sufficient to prove that the function $\vartheta_{00}$ evaluated at 
$
(\tau,z)=(\frac{i c}{\pi},\frac{b}{2\pi})
$
satisfies the ODE (\ref{ODE}).
The differential operator defined by the left hand side of  (\ref{ODE}), say $\mathcal{D}$,
has the following two properties: for
$
\left(
\begin{smallmatrix}
\alpha & \beta \\ \gamma & \delta
\end{smallmatrix}
\right)
$
in a suitable subgroup of $\SL(2,\RR)$ and $(\lambda,\mu)$ in a suitable sublattice
of $\RR^2$, assume that a function $h$ transforms as follows
\begin{equation}\label{JacobiForm}
h\left(\frac{b}{\gamma c+\delta},\frac{\alpha c+\beta}{\gamma c+\delta}\right)=
(\gamma c+\delta)^{\frac{1}{2}}
e^{\frac{\gamma b^2}{\gamma c+\delta}}h(b,c),
\quad
h(b+\lambda c+\mu,c)=e^{(\lambda^2 c+2\lambda b)}h(b,c)
\end{equation}
then
\begin{align*}
\mathcal{D}h\left(\frac{b}{\gamma c+\delta},\frac{\alpha c+\beta}{\gamma c+\delta}\right)&=
(\gamma c+\delta)^{12}
e^{6\frac{\gamma b^2}{\gamma c+\delta}}\mathcal{D}h(b,c),\\
\mathcal{D}h(b+\lambda c+\mu,c)&=
e^{6(\lambda^2 c+2\lambda b)}\mathcal{D}h(b,c).
\end{align*}
This indicates that the operator $\mathcal{D}$ sends, up to the condition on
the Fourier expansion, Jacobi forms of weight one-half
and index one-half to Jacobi forms of weight $12$ and index $3$ on a suitable subgroup of 
$\SL(2,\RR)$. Recalling that the Jacobi theta series $\vartheta_{00}$ can be viewed as a 
Jacobi form of weight one-half and index one-half (with a multiplier system)
on the principal congruence subgroup of level 4, checking that the function $\vartheta_{00}$ evaluated at 
$
(\tau,z)=(\frac{i c}{\pi},\frac{b}{2\pi})
$
satisfies the ODE (\ref{ODE}) reduces to check it for sufficiently many Fourier coefficients 
(finite dimension of spaces of Jacobi forms) which can be done by any computer algebra systems.
We conclude that any Jacobi theta series evaluated at 
$
(\tau,z)=(\frac{i c}{\pi},\frac{b}{2\pi})
$
satisfies the ODE (\ref{ODE}) and the PDE 
$
\psi_c=\psi_{bb}.
$
Taking for example $\psi(b,c)=\vartheta_{00}(\frac{i c}{\pi},\frac{b}{2\pi})$, we get for the function $f$
\[
f(b,c)=2\frac{\psi_b}{\psi}(b,c)=-4
(
\sin(b)e^{-c}-\sin(2b)e^{-2c}+(\sin(b)+\sin(3b))e^{-3c}-\sin(4b)e^{-4c}+\ldots
).
\]

\subsection{Equations of the form $u_{x_1x_2}+u_{x_1x_3}u_{x_2x_3}r(u_{x_3x_3})=0$}\

This case is obtained as a specialisation of the previous one: 
\[
f(u_{x_2x_3}, u_{x_3x_3})=u_{x_2x_3}r(u_{x_3x_3}).
\]

The integrability conditions in this case are obtained by substituting $f(b,c)=b\, r(c)$ in the system (\ref{SPDE}). 
This leads to a single third-order ordinary differential equation for the function $r$, we set 
$
r^{(k)}=\frac{d^k r}{dc^k}
$:
\begin{equation}\label{r1}
r'''(r'-r^2)-(r'')^2+4r^3\, r''+2(r')^3-6(r r')^2 = 0.
\end{equation}
This case has already been investigated in section 4.5 of \cite{BFT}.  Let us briefly recall the results
obtained there: we denote by $\mathcal{D}r(c)$ the left hand side of (\ref{r1}).
As noticed in \cite{BFT}, the expression $\mathcal{D}r(c)$ has a $\SL(2,\RR)$-invariance:
for any 
$
\left(
\begin{smallmatrix}
\alpha & \beta \\ \gamma & \delta
\end{smallmatrix}
\right)
\in \SL(2,\RR)
$, 
assume that the function $r$ satisfies the following functional equation 

\begin{equation}\label{act}
\tilde r(c)=r\left(\frac{\alpha c+\beta}{\gamma c+\delta}\right)=(\gamma c+\delta)^{2}r(c)+\gamma(\gamma c+\delta)
\end{equation}
then we have
\[
\mathcal{D}\tilde r(c)=\mathcal{D}r\left(\frac{\alpha c+\beta}{\gamma c+\delta}\right)=(\gamma c+\delta)^{12}\mathcal{D}r(c).
\]
Then modulo the $\SL(2,\RR)$-action given by (\ref{act}), the generic solution of (\ref{r1}) is given by ($c<0$): 
\[
r(c)=1-8\sum_{n\geq 1}\sigma_1^{-}(n) e^{4nc}
\quad
\text{with} 
\quad
\sigma_1^{-}(n)=\sum_{d\vert n} (-1)^dd.
\]
Note that for $\tau\in \mathfrak{H}_1$, 
$
r(\frac{\pi i \tau}{2})=1-8\sum_{n\geq 1}\sigma_1^{-}(n) e^{2\pi in\tau}
$
is a quasi-modular form of weight 2 on the congruence subgroup $\Gamma_0(2)$.
In fact we have 
$
r(\frac{\pi i \tau}{2})=(4e_2(2\tau)-e_2(\tau))/3
$
where $e_2$ is the Eisenstein series of weight 2 on $\SL(2,\ZZ)$, see (\ref{Eis2}).

We would like to give another description of the generic solution which will make a connection with the Legendre
family of elliptic curves. We make the change of functions $r=g'/g$ in (\ref{r1}), then the function $g$ satisfies the following
fourth-order ordinary differential equation:
\begin{equation}\label{g}
g''''(g''g-2g'^2)g^3-g'''^2g^4+2g^2g'g'''(4g'^2+g''g)-g''^3g^3-3g''^2g'^2g^2-12g''g'^4g+8g'^6=0.
\end{equation} 
Section 5 of \cite{CO} proposed a method for linearising any equation of the form  (\ref{r1}) which has been carried out 
explicitly along the proof of theorem 3 of \cite{FO}. This gives the following hypergeometric differential equation associated 
to the equation (\ref{r1}): 
\[
s(1-s)w''(s)+(1-2s)w'(s)-\frac{1}{4}w(s)=0.
\] 
Let $w_1$ and $w_2$ be two linearly independent solutions  of this  hypergeometric equation 
with the Wronskian normalised as  $w_2w_{1, s}-w_1w_{2, s}=\frac{1}{s(1-s)}$.
These solutions can be written as periods of the holomorphic differential 1-form $\omega$ given by
\[
\omega=\frac{dt}{\sqrt{t(t-1)(t-s)}}
\]
of the elliptic curve $v^2=t(t-1)(t-s)$, we assume $s\in (0, 1)$. 
Let us choose a basis of solutions such that $w_1$ has a logarithmic singularity at zero 
while $w_2$ is the standard hypergeometric series with parameters $(1/2,1/2;1)$:
\begin{align*}
w_1(s)&=-{i} \int_{-\infty}^{0}\omega+\frac{4\ln 2}{\pi}\int_1^{\infty}\omega=
w_2(s)\ln s+2\sum_{n=1}^{\infty}\left(\frac{(2n-1)!!}{2^nn!}\right)^2\left(\sum_{k=1}^n\frac{1}{k(2k-1)}\right)s^n,\\
w_2(s)&=\frac{1}{\pi} \int_1^{\infty}\omega=F\left(\frac{1}{2}, \frac{1}{2}; 1; s\right).
\end{align*}
This choice satisfies the Wronskian constraint. Noticing that  equation (\ref{g}) has 
the following $\SL(2,\RR)$-symmetry which comes directly from (\ref{act}):
\[
\tilde c=\frac{\alpha c+ \beta }{\gamma c+ \delta},
\quad 
\tilde g= (\gamma c+ \delta) g,  
\] 
we obtain a generic solution via parametric formulae (viewing $s$ as a parameter)

\begin{equation}\label{par}
c=\frac{w_1(s)}{w_2(s)}
\quad
\text{and}
\quad
g=w_2(s).
\end{equation}
It is a classical fact that the first relation can be  inverted and this gives
$
s(c)=\frac{\vartheta_{10}^4(ic,0)}{\vartheta_{00}^4(ic,0)}
$
and the substitution into $g=w_2$ gives
\[
g(c)=w_2(s(c))=F\left(\frac{1}{2}, \frac{1}{2};1; s(c)\right)=F\left(\frac{1}{2}, \frac{1}{2};1;\frac{\vartheta_{10}^4(ic,0)}{\vartheta_{00}^4(ic,0)}\right)=
\vartheta_{00}^2(ic,0).
\]
We refer to \cite{BGHZ, Hu} for a review of these classical formulae.

\begin{remark}
Since the action of the group $\SL(2,\ZZ)$ permutes the Jacobi theta series with characteristics in 
$\left\{0,1\right\}$ and the equation (\ref{g}) admits a $\SL(2,\RR)$-symmetry,
another solution to the equation (\ref{g}) can be chosen as $g(c)=\vartheta_{10}^2(ic,0)$. For 
$\tau\in \mathfrak{H}_1$, let $f(\tau)=\vartheta_{10}^2(\tau,0)$. It is a well-known
fact that the function $f$ is modular form of weight 1 on $\Gamma_0(2)$ with a character.
A direct computation shows that the logarithmic derivative of $f$ transforms like a quasi-modular form of 
weight 2 on $\Gamma_0(2)$, note the cancellation of the character, and we have
\[
\frac{f'}{f}(\tau)=2\frac{\vartheta_{10}'(\tau,0)}{\vartheta_{10}(\tau,0)}=\frac{\pi i}{2}\, r(\frac{\pi i \tau}{2}).
\]
The latter formula connects the two descriptions presented in this section.
\end{remark}

\section{\textbf{Concluding remarks}}
The attempts to generalise/specialise Theorem \ref{main} lead to the following observations:

\begin{itemize}

\item Some compactifications of the hyperelliptic divisor in genus 3 \cite{FSM} should be related to the classification of 
special integrable Hirota type equations (\ref{main}) that have {\it continuous} symmetries from the 
equivalence group $\Sp(6,\RR)$. 
Such degenerations  could be interesting from the point of view of their potential applications. 

\item The construction of Odesskii and Sokolov \cite{Odesskii2} can be adapted to show that for any 
higher genus $g \geq 4$ the hyperelliptic locus defines a 3D integrable hierarchy of Hirota-type equations
(in the notation of \cite{Odesskii2} this corresponds to the choice of  constants $
s_1=\dots =s_g=s_{2g}=s_{2g+1}=\frac{1}{2}, \ s_{g+1}=s_{2g-1}=-\frac{1}{2}, \ n=2g-1, \ k=g$). 
Here the equations of the hierarchy coincide with the defining equations of the hyperelliptic locus 
\cite{Mumford, Poor,  SM, Gru}, which is known to be characterised by the vanishing of  $\frac{1}{2}(g-1)(g-2)$ 
theta constants (Hirota equations result on the substitution of $\tau=i \text{Hess}(u)$ for the $g\times g$ period matrix).

\item  Although it would be tempting to conjecture that the Schottky divisor (image of the Torelli map 
${\mathcal M}_4\hookrightarrow {\mathcal A}_4$ 
from the moduli space of curves of genus 4 to the moduli space of principally polarised Abelian varieties \cite{Donagi}) 
corresponds to an integrable  Hirota type equation in 4D, this is no longer the case. Recent results of \cite{FKN} show that in 4D the requirement of integrablity implies the symplectic Monge-Amp\`ere property, which leads to a complete list of integrable heavenly-type equations classified in \cite{DF}. Thus, the occurrence of modular forms and theta functions in the classification of integrable Hirota type equations is the essentially 3-dimensional phenomenon. 

\item Integrable symplectic Monge-Amp\`ere equations in 4D correspond to hypersurfaces $M^9$ in the Lagrangian Grassmannian $\Lambda^{10}$ which can be characterised as special hyperplane sections of the Pl\"ucker embedding
$\Lambda^{10}\subset \mathbb{P}^{41}$ (see \cite{DF}). In 3D we have a similar situation: embedding the quotient of the Siegel 
upper half space $\mathfrak{H}_3$ by the group $\Gamma_3(4, 8)$ into $\mathbb{P}^{35}$ via the 36 even theta constants we obtain a projective variety
cut out by quartics (see \cite{Glass}). In this picture, hyperelliptic divisor is cut out by the coordinate hyperplanes
in $\mathbb{P}^{35}$. 

\item Finally, it would be interesting to obtain a purely computational proof of theorem \ref{T1} by directly proving that even  theta constants satisfy the integrability conditions obtained by the method of hydrodynamic reductions, see section \ref{sec:HR}.

\end{itemize}

\section*{\textbf{Acknowledgements}}
We thank B. Dubrovin, I. Krichever,  M. Pavlov, A. Prendergast-Smith and A. Veselov for clarifying discussions. 
This research was  supported by the EPSRC grant EP/N031369/1.

\end{document}